\documentclass[11pt]{article}
\usepackage{graphicx}

\newcommand{\rsdecay}{\mbox{\ensuremath{\Dz \to K^{-} \pi^{+} \pi^{+}\pi^{-}}}}
\newcommand{\wsdecay}{\mbox{\ensuremath{\Dz \to K^{+} \pi^{-} \pi^{+}\pi^{-}}}}
\newcommand{\wsdecayb}{\mbox{\ensuremath{\Dzb \to K^{-} \pi^{+} \pi^{-}\pi^{+}}}}
\newcommand{\dm}{\ensuremath{\Delta m}}

\newcommand{\mKppp}{\ensuremath{m_{K\pi\pi\pi}}}
\newcommand{\tKppp}{\ensuremath{t_{K\pi\pi\pi}}}
\newcommand{\rightsign}{RS}
\newcommand{\wrongsign}{WS}

\newcommand{\BABARPubYear}    {06}
\newcommand{\BABARConfNumber} {031}
\newcommand{\SLACPubNumber} {12019}

\input pubboard/babarsym
 
\long\def\inst#1{\par\nobreak\kern 4pt\nobreak
    {\it #1}\par\vskip 10pt plus 3pt minus 3pt}

\begin{document}
{\pagestyle{empty}

\begin{flushright}
\babar-CONF-\BABARPubYear/\BABARConfNumber \\
SLAC-PUB-\SLACPubNumber \\
\end{flushright}

\par\vskip 5cm

\begin{center}
\Large \bf Search for \Dz-\Dzb\ mixing in the decays \wsdecay

\end{center}
\bigskip

\begin{center}
\large The \babar\ Collaboration\\
\mbox{ }\\
\today
\end{center}
\bigskip \bigskip

\begin{center}
\large \bf Abstract

\end{center}
We present a search for \Dz-\Dzb\ mixing in the decays
\wsdecay\ using 230.4\invfb of data collected with the \babar\
detector at the \pep2 $e^+ e^-$ collider at SLAC.
Assuming \CP\ conservation, we measure the time-integrated mixing rate
$R_M = (0.019\ \mbox{}^{+0.016}_{-0.015}\,(\textrm{stat.}) \pm 0.002\,(\textrm{syst.})) \%$,
and $R_M < 0.048\%$ at the 95\% confidence level.  Using a frequentist method,
we estimate that the data are consistent with no mixing at the 4.3\% confidence level.
We present results both with and without the assumption of \CP\ conservation.
By combining the value of $R_M$ from this analysis with that obtained from
an analysis of the decays $\Dz \to K^+ \pi^- \pi^0$, we find
$R_M = (0.020\,\mbox{}^{+0.011}_{-0.010})\%$,
where the uncertainty is statistical only. 
We determine the upper limit $R_M < 0.042\%$ at the 95\% confidence level,
and we find the combined data are consistent with the no-mixing
hypothesis at the 2.1\% confidence level.  
\vfill
\begin{center}

Submitted to the 33$^{\rm rd}$ International Conference on High-Energy Physics, ICHEP 06,\\
26 July---2 August 2006, Moscow, Russia.

\end{center}

\vspace{1.0cm}
\begin{center}
{\em Stanford Linear Accelerator Center, Stanford University, 
Stanford, CA 94309} \\ \vspace{0.1cm}\hrule\vspace{0.1cm}
Work supported in part by Department of Energy contract DE-AC03-76SF00515.
\end{center}

\newpage
} 

\begin{center}
\small

The \babar\ Collaboration,
\bigskip

%
{B.~Aubert,}
{R.~Barate,}
{M.~Bona,}
{D.~Boutigny,}
{F.~Couderc,}
{Y.~Karyotakis,}
{J.~P.~Lees,}
{V.~Poireau,}
{V.~Tisserand,}
{A.~Zghiche}
\inst{Laboratoire de Physique des Particules, IN2P3/CNRS et Universit\'e de Savoie,
 F-74941 Annecy-Le-Vieux, France }
% ...

\end{center}\newpage

\section{INTRODUCTION}
\label{sec:Introduction}
Transitions between the flavor eigenstates $|\Dz\rangle$ and $|\Dzb\rangle$
are called $D$ mixing, which is expected to have a very small rate in the
Standard Model. Due to significant contributions from long-distance
effects~\cite{DMixTheory}, an accurate estimate is difficult to obtain,
but typical theoretical estimates of the time-integrated mixing rate are
$R_M \sim \mathcal{O}(10^{-6}\textrm{--}10^{-4})$.
The \babar\ collaboration has previously reported searches for $D$ mixing
in the decays to \CP-even eigenstates~\cite{Aubert:2003pz},
in the decay $\Dz \to K^+ \pi^-$~\cite{Aubert:2003ae},
and in semileptonic decays~\cite{Aubert:2004bn}.
A recent analysis of the decay $\Dz \to K^+ \pi^- \pi^0$
set an upper limit $R_M < 0.054\%$ at the 95\% confidence level
with a data sample consistent with no mixing at the 4.5\%
confidence level~\cite{Wilson:2006sj}.
The most stringent constraints on $D$-mixing parameters to
date have been obtained by analyzing the decay
$\Dz \to K^+\pi^-$~\cite{Zhang:2006dp}; the rate
is determined to be $R_M < 0.040\%$ at the 95\% confidence level.

We search for the process
$|\Dz\rangle\to|\Dzb\rangle$
by analyzing the decay of a particle
known to be created as a $|\Dz\rangle$~\cite{conjugates}.  We distinguish doubly
Cabibbo-suppressed (DCS) contributions from Cabibbo-favored (CF) mixed contributions
by the decay-time distribution in the reconstructed wrong-sign
(\wrongsign) decay \wsdecay.
The right-sign (RS) decay \rsdecay\ is a normalization mode in this analysis.

The two mass eigenstates
\begin{equation}
\label{eq:massstates}
 |D_{A,B}\rangle = p|\Dz\rangle \pm q|\Dzb\rangle
\end{equation}
generated by mixing dynamics have different
masses $(m_{A,B})$ and widths $(\Gamma_{A,B})$, with $|p/q|=1$ if \CP\
is conserved in mixing. We parameterize the mixing process with the quantities
\begin{equation}
x \equiv 2\frac{m_{B} - m_{A}}{\Gamma_{B} + \Gamma_{A}},\ \ \ \ \ %
y \equiv \frac{\Gamma_{B} - \Gamma_{A}}{\Gamma_{B} + \Gamma_{A}}%
\textrm{.}
\end{equation}
For a multibody \wrongsign\ decay, the time-dependent decay
rate, relative to a corresponding \rightsign\ rate, is
approximated by~\cite{Blaylock:1995ay}
\begin{eqnarray}
\label{eq:tdratemult}
 & {\displaystyle\frac{\Gamma_{\textrm{WS}}(t)}{\Gamma_{\textrm{RS}}(t)} =%
   \tilde{R}_D + \alpha\tilde{y}'\sqrt{\tilde{R}_D}\,(\Gamma t)%
   + \frac{\tilde{x}'^2+\tilde{y}'^2}{4}(\Gamma t)^2 } & \\
 &  0 \leq \alpha \leq 1\textrm{,} & \nonumber
\end{eqnarray}
where the tilde indicates quantities that have been integrated over the selected
phase-space regions.  Here, $\tilde{R}_D$ is the integrated DCS branching ratio;
$\tilde{y}' = y\cos\tilde{\delta} - x\sin\tilde{\delta}$ and
$\tilde{x}' = x\cos\tilde{\delta} + y\sin\tilde{\delta}$, where
$\tilde{\delta}$ is an unknown integrated strong-phase difference;
$\alpha$ is a suppression factor that accounts for
strong-phase variation over the region; and $\Gamma$ is the
mean width.  The
time-integrated mixing rate
$R_M = (\tilde{x}'^2+\tilde{y}'^2)/2 = (x^2+y^2)/2$ is independent of decay mode
and should be consistent among mixing measurements.
Additionally, while the branching ratio of DCS to CF decays depends on position
in the Dalitz plot, the mixing rate does not.

We also search for \CP\ violation in a mixing signal by fitting to the \wsdecay\ and
\wsdecayb\ samples separately. We consider \CP\ violation
in the interference between the DCS and mixed contributions, parameterized
by an integrated \CP-violating--phase $\tilde{\phi}$, as well as \CP\ violation
in mixing, parameterized by $|p/q|$.
We assume \CP\ invariance in both the DCS and CF decay rates.  The substitutions
\begin{eqnarray}
 & \displaystyle \alpha\tilde{y}' \to \left|\frac{p}{q}\right|^{\pm 1}%
 (\alpha\tilde{y}'\cos\tilde{\phi} \pm \beta\tilde{x}'\sin\tilde{\phi}) & \\
 & \displaystyle (x^2+y^2) \to \left|\frac{p}{q}\right|^{\pm 2}(x^2+y^2) &
\end{eqnarray}
are applied to Equation~\ref{eq:tdratemult},
using $(+)$ for $\Gamma(\wsdecayb)/\Gamma(\rsdecay)$ and $(-)$ for the
charge-conjugate ratio.  The parameter $\beta$ is analogous to $\alpha$
and accounts for net $\phi$ variation.

\section{THE \babar\ DETECTOR AND DATASET}
\label{sec:babar}
We use 230.4\invfb of data collected with the \babar\ detector~\cite{Aubert:2001tu}
at the \pep2 $e^+ e^-$ collider at SLAC.
Charged particles are detected
and their momenta measured by a combination of a cylindrical drift chamber (DCH)
and a silicon vertex tracker (SVT), both operating within a
1.5\,T solenoidal magnetic field.
A ring-imaging Cherenkov detector (DIRC) is used for
charged-particle identification. Photon energies are measured with a
CsI electromagnetic calorimeter (EMC).
We use information from the DIRC and energy-loss measurements in the
SVT and DCH to identify charged-kaon and -pion candidates.
The data set includes $e^+e^-$ collisions at
and 40\mev below the \FourS\ resonance.  All selection criteria were finalized
before searching for evidence of mixing in the data.

\section{ANALYSIS METHOD}
\label{sec:Analysis}

We reconstruct the decays $\Dstarp \to \Dz\pi_s^{+}$ and the charge of the
soft pion, $\pi_s^{\pm}$, is used to determine the flavor of the \Dz\
candidate. In order to obtain a pure data sample,
selection of \Dz\ candidates includes a requirement of
center-of-mass momentum greater than 2.4\gevc and the application
of strict particle-identification (PID) requirements to the daughters
of the \Dz.  We accept decays with an invariant mass
$1.815 < \mKppp < 1.915\gevcc$ and 
an invariant mass difference 
$0.1396 < \dm < 0.1516\gevcc$, where
$\dm \equiv m_{K\pi\pi\pi\pi_{s}} - \mKppp$. 
We also require
that neither $\pi^+\pi^-$ combination of candidate \Dz\ daughters
have an invariant mass within
$20\,\mevcc$ of the $K^0_S$ value given in the
Review of Particle Physics (RPP)~\cite{Eidelman:2004wy}. This cut suppresses
background from the singly Cabibbo-suppressed decay $\Dz\to
K^+\Kzb\pi^-$ followed by $\Kzb\to \pi^+\pi^-$.

The candidate masses and decay times are derived from
a vertex fit.
First, the \Dz\ and \Dstarp\ decay vertices are determined
in separate geometric fits, and the $\chi^2$ probability
of each fit is required to be greater than 0.005.
The candidate \Dstarp-decay tree is then fit for simultaneously
optimal \Dstarp\ and \Dz\ decay vertices~\cite{Hulsbergen:2005pu} with
the \Dstarp\ decay vertex constrained to the beamspot region.
We select events for which the $\chi^2$ probability of this fit
is greater than $0.01$.
From this fit, a \Dz\ decay time, \tKppp, and
uncertainty, $\sigma_t$, are calculated using the three-dimensional
flight path.  The full covariance
matrix, including correlations between the two vertices, is
used in the $\sigma_t$ estimate.  For signal events, the mean
$\sigma_t$ is near 0.29\ps; we accept decays with
$\sigma_t < 0.5\ps$.
The world-average \Dz\ lifetime
is 0.41\ps~\cite{Eidelman:2004wy}.

To separate correctly reconstructed decays from background,
and to distinguish mixing contributions from DCS contributions,
unbinned extended maximum likelihood fits to the data sample are performed.
Probability density functions (PDFs) are fit in two stages to
the distributions $(\mKppp,\dm,\tKppp)$.  First,
the $(\mKppp,\dm)$ plane is considered to discriminate
between signal and background; optimal
PDF parameters are established in these dimensions.
Second, a fit to $\tKppp$ is performed, retaining the PDF-shape parameters
of the previous fit to construct a three-dimensional likelihood $\mathcal{L}$.

\begin{figure}
\begin{center}

\includegraphics[width=\linewidth]{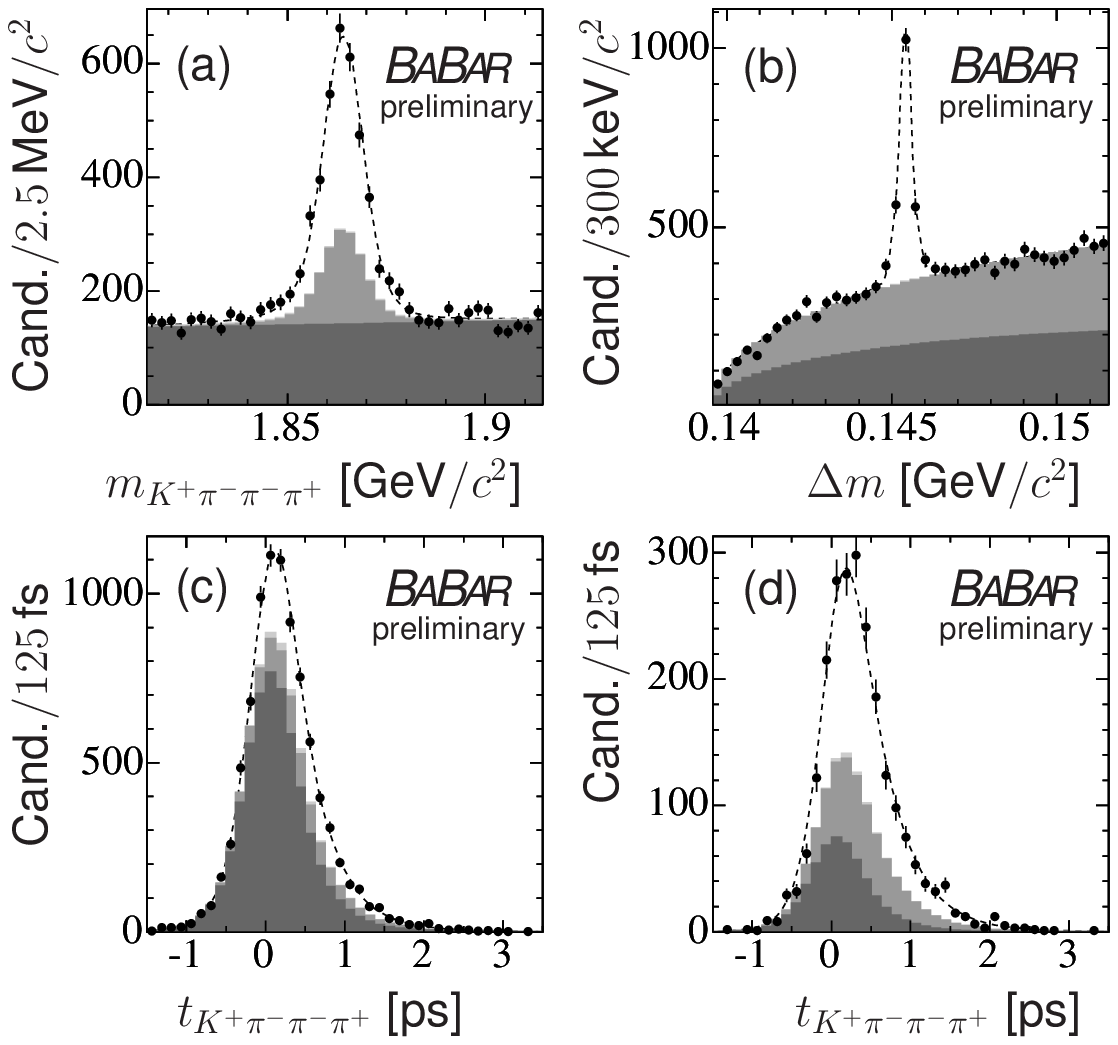}
\caption
{
Distributions of \wrongsign\ data with fitted PDFs
(described in Sec.~\ref{sec:Analysis}) overlaid.
The \mKppp\ distribution (a) requires
$0.14487 < \dm < 0.14587\gevcc$; 
the \dm\ distribution (b) requires
$1.859 < \mKppp < 1.869\gevcc$; and the \tKppp\ distribution (c)
requires \dm\ cuts as the same as (a). The \tKppp\ distribution 
(d) requires both cuts from (a) and (b). 
In each of the above histograms, the white area beneath the
dotted line represents signal events,
the light gray represents swapped $K^\pm\pi^\mp$,
the medium gray represents misassociated $\pi_s^{\pm}$ events, and
the dark gray represents remaining combinatorial background.
}
\label{fig:datafit}
\end{center}
\end{figure}

The signal yields from the fit to the $(\mKppp,\dm)$ plane are
listed in Table~\ref{tbl:candnumbers}.  A simultaneous fit is performed
to both the large sample of \rightsign\ decays and the relatively small
sample of \wrongsign\ decays; thus, signal shape parameters associated
with the \wrongsign\ sample are precisely determined by the \rightsign\ sample,
and all associated systematic uncertainties are suppressed.
The fit to these distributions is shown for the \wrongsign\ sample in
Figure~\ref{fig:datafit}(a,b).

\begin{table}[!htp]
\caption{
Signal-candidate yields determined by the
two-dimensional fit to the $(\mKppp,\dm)$
distributions
for the \wrongsign\
and \rightsign\ samples.  
Uncertainties are those calculated from the fit.}
\begin{center}
\begin{tabular}{lrr}
\hline
        & \Dz\ Cand.                        &   \Dzb\ Cand.\\
\hline
 WS     & $(1.162  \pm 0.053) \times 10^3$  &   $(1.040  \pm 0.051) \times 10^3$ \\
 RS     & $(3.511  \pm 0.006) \times 10^5$  &   $(3.492  \pm 0.006) \times 10^5$ \\
 \hline
\end{tabular}
\label{tbl:candnumbers}
\end{center}
\end{table}

The sources of background remaining in the sample
may be characterized by three
categories in the likelihood fits to data.
The background that peaks in the \mKppp\ distribution
is due to correctly reconstructed \Dz\ decays with a
misassociated $\pi_s^{+}$; this category has the decay-time
distribution of the \rightsign\ signal.
Second, remaining combinatorial background is present as a
nonpeaking component of both distributions.
This distribution is empirically described by a
Gaussian with a power-law tail.
The third category is due to correctly reconstructed \Dstarp
decays with a misreconstructed \Dz, for which the kaon and a pion have been mistaken for
each other. This category has the signal lifetime distribution.
A three-dimensional likelihood is maximized in a fit to $\tKppp$, after the shape
parameters are determined in the two mass distributions.

The \rightsign\ PDF is fit to the $\tKppp$ distribution to determine the \Dz\ lifetime and
the detector-resolution parameters.
The signal shape of \tKppp\ is an exponential function
convolved with a double-Gaussian resolution function.  The Gaussians
have different widths and means;  the width of each
Gaussian is a scale factor multiplied by $\sigma_t$, which
is determined for each event.  The two different scale factors are determined by
the fit to the data.  We find a $\Dz$ lifetime
consistent with the nominal value.

The \wrongsign\ signal PDF in \tKppp\ is
a function based on Equation~\ref{eq:tdratemult} convolved
with the double Gaussians described above.  The $\Dz$ lifetime
and resolution scale factors and means, determined by the fit to the
\rightsign\ \tKppp\ distribution,
are fixed.  We fit the \wrongsign\
PDF to the \tKppp\ distribution allowing yields and
background shape parameters to vary.  The fit to the
\tKppp\ distribution is shown for the \wrongsign\ sample in
Figure~\ref{fig:datafit}(c,d).

\section{SYSTEMATIC STUDIES}
\label{sec:Systematics}
We quantify systematic uncertainties by performing the analysis with
the following changes, in order of decreasing significance: 
the selection of events based on $\sigma_t$,
the decay-time resolution function,
the background PDF shape in the \mKppp\ distribution,
and the measured \Dz\ lifetime value. 
The selection requirement of $\sigma_t$ may skew the
decay-time distribution if there is a correlation between
the two distributions; it is investigated
by moving the selection criterion from 0.5\ps\ to 0.6\ps.
The double-Gaussian resolution function is investigated by
fixing one of the two scale factors to unity, determining
the other factor from a fit to the \rightsign\ data, and performing
the decay-time fit to the \wrongsign\ data.
The PDF used to describe the background contribution in the
\mKppp\ distribution is changed from an exponential to
a second-order polynomial. This change allows some fraction of events
to be weighted toward background, and so affects the
number of events contributing to the mixing signal.
Finally, the fitted lifetime from the decay-time fit
to the \rightsign\ data is not as accurate as the
value listed in the RPP~\cite{Eidelman:2004wy};
this systematic uncertainty is estimated by setting the lifetime
to the value given in the RPP.
The combined systematic uncertainties for most quantities
in Table~\ref{tbl:results}
are smaller than statistical uncertainties
by a factor of 5;  the systematic uncertainties on
$(\alpha \tilde{y}'\cos \tilde{\phi})$ and
$(\beta \tilde{x}'\sin \tilde{\phi})$ do not account for
correlated uncertainties between the \Dz\ and \Dzb\
samples, and thus are conservatively estimated.

\section{RESULTS}

\begin{table}[!htp]
\caption
{
Mixing results assuming \CP\ conservation
(\Dz\ and \Dzb\ samples are not separated) and
manifestly permitting \CP\ violation (\Dz\ and \Dzb\ samples are
fit separately).  The first listed uncertainty is
statistical, and the second is systematic.
}
\begin{center}
\begin{tabular}{cr|cr}
\hline
 & \multicolumn{1}{c}{\CP\ conserved} &
 & \multicolumn{1}{c}{\CP violation allowed} \\
\hline
 \rule{0em}{3ex}$R_M$ &
\multicolumn{1}{r}{$(0.019\ \mbox{}^{+0.016}_{-0.015} \pm 0.002) \%$} &
 &
\multicolumn{1}{r}{$(0.017\ \mbox{}^{+0.017}_{-0.016} \pm 0.003) \%$} \\
\multicolumn{1}{r}{} \\
\hline
 \rule{0em}{4.25ex}\rule[-3ex]{0em}{3ex}$\alpha \tilde{y}'$ &
 $-0.006\ \mbox{}^{+0.005}_{-0.005} \pm 0.001$ &
 \parbox{2em}{%
 \begin{tabular}{l}
 $\alpha \tilde{y}'\cos \tilde{\phi}$ \\
 \rule{0em}{2.75ex} \\
 \rule{0em}{3.25ex}$\beta \tilde{x}'\sin \tilde{\phi}$ \\
 \rule{0em}{2.75ex}
 \end{tabular}} &
 \parbox{10em}{%
 \begin{tabular}{r}
 \\
 \rule{0em}{2.75ex}$-0.006\ \mbox{}^{+0.008}_{-0.006} \pm 0.006 $ \\
 \rule{0em}{3.25ex} \\
 \rule{0em}{2.75ex}$0.002\ \mbox{}^{+0.005}_{-0.003} \pm 0.006$
 \end{tabular}} \\
\hline
 \rule{0em}{3ex}& &
 \rule[-1.25ex]{0em}{1.25ex}$|p/q|$ & $1.1\ \mbox{}^{+4.0}_{-0.6} \pm 0.1 $\\
\hline
\end{tabular}
\label{tbl:results}
\end{center}
\end{table}

\begin{figure}
\begin{center}

\begin{tabular}{p{0.48\linewidth}p{0.48\linewidth}}
\includegraphics[width=\linewidth]{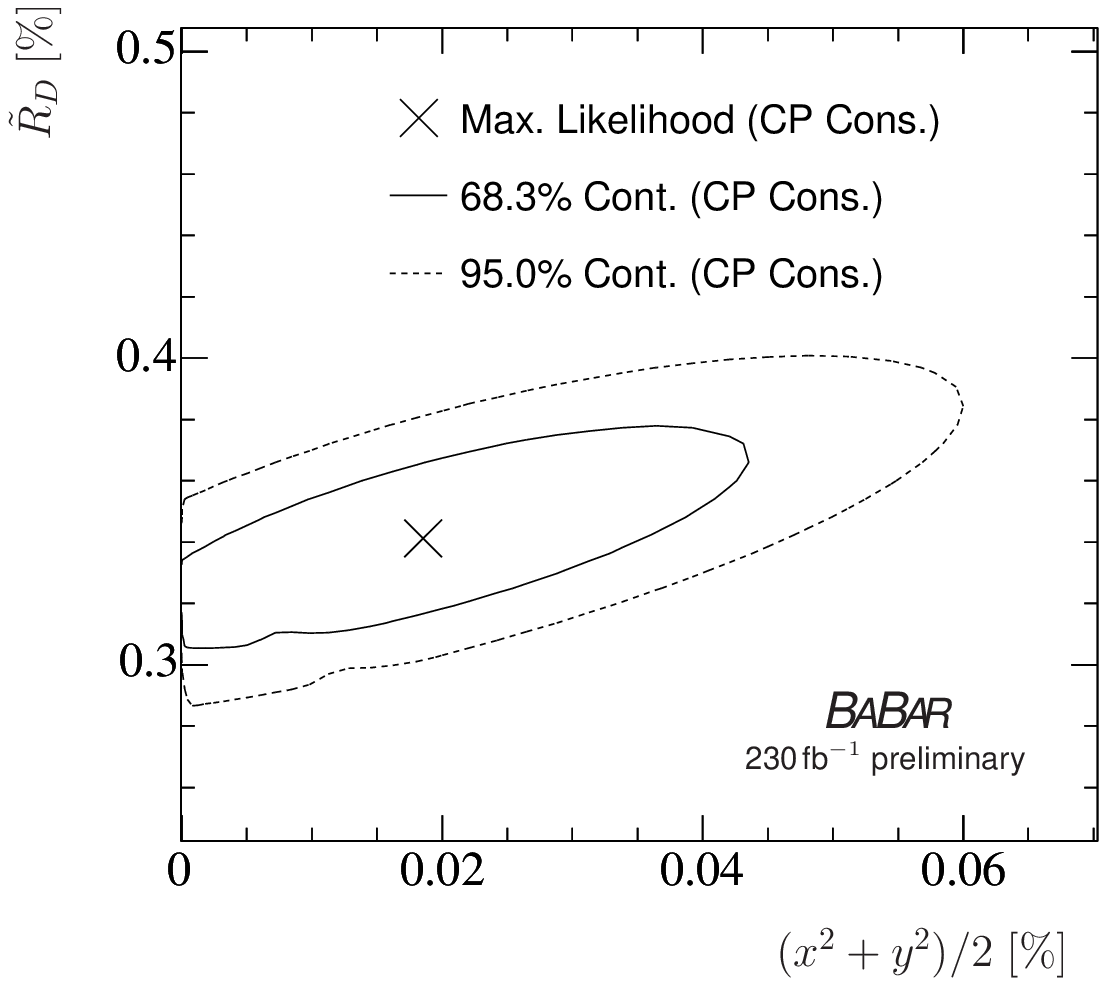}
&
\includegraphics[width=\linewidth]{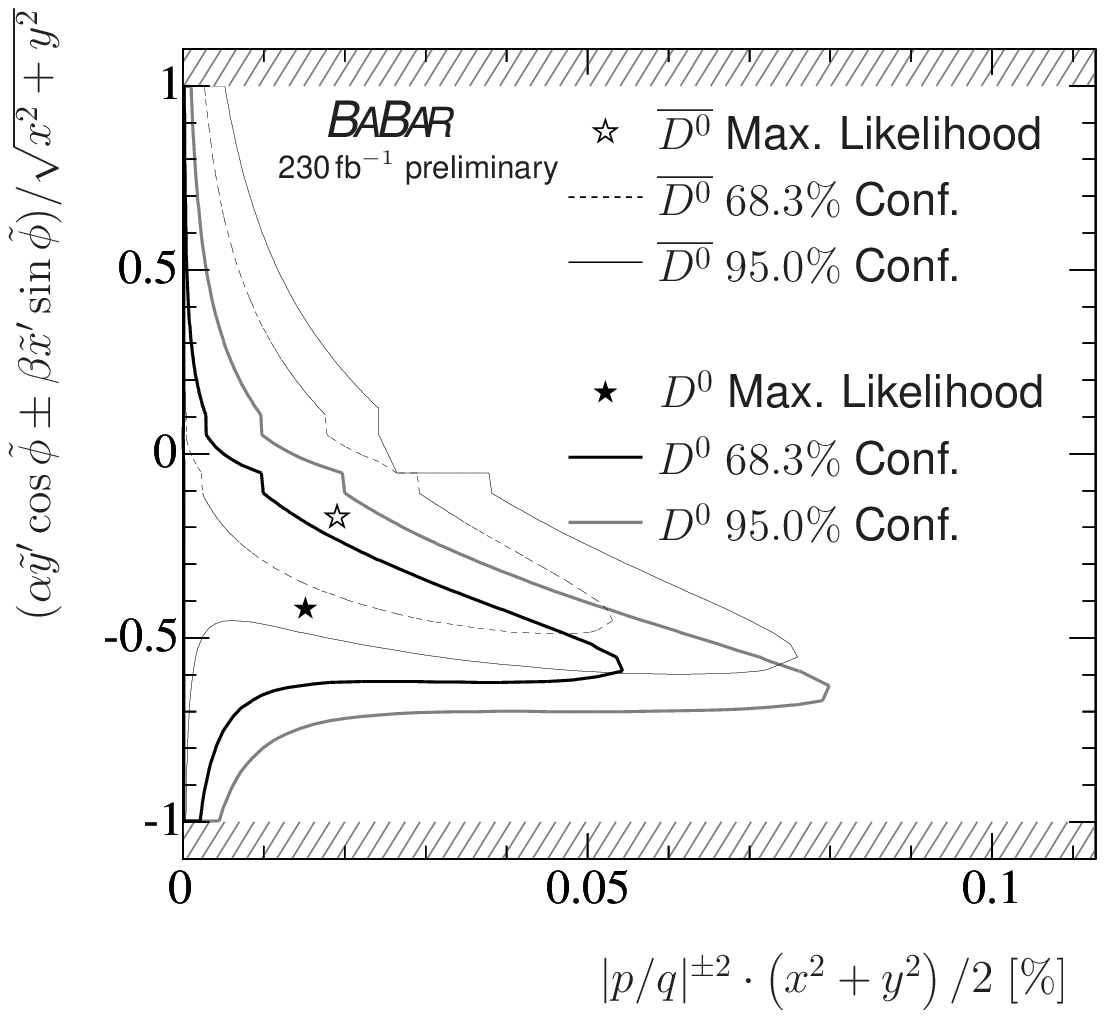}
\end{tabular}
\caption{
Left: Contours of constant $\Delta\ln\mathcal{L}=1.15,3.0$ in terms of
the doubly Cabibbo-suppressed branching ratio and the time-integrated
mixing rate.  The upward slope of the contour indicates
negative interference between DCS and mixed contributions.
Right: Contours of constant $\Delta\ln\mathcal{L}=1.15,3.0$ in terms of
the normalized interference term and the integrated
mixing rate, for the \Dz\ and \Dzb\ samples separately.
The hatched regions are physically forbidden.
}
\label{fig:contour_rmix}
\end{center}
\end{figure}

The results of the decay-time fit, both with and 
without the assumption of \CP\ conservation in a mixing signal,
are listed in
Table~\ref{tbl:results}.  The statistical uncertainty
of a particular parameter is obtained by finding its extrema
for $\Delta\ln\mathcal{L}=0.5$; in finding
the extrema, the likelihood is kept maximal by
refitting the remaining parameters.
Contours of constant $\Delta\ln\mathcal{L}=1.15,3.0$,
enclosing two-dimensional coverage probabilities of
68.3\% and 95.0\%, respectively,
are shown in Figure~\ref{fig:contour_rmix}.

We note that $\Delta\ln\mathcal{L}$ as a function of
the quantity $\textrm{sign}(\alpha \tilde{y}')\times R_M$ is
approximately parabolic.  The two-sided interval
$-0.048\% < \textrm{sign}(\alpha \tilde{y}')\times R_M < 0.048\%$
contains 95\% coverage probability; thus, we quote $R_M < 0.048\%$ as our upper 
limit on the integrated mixing rate under the assumption of \CP\ conservation.

A feature of $\Delta\ln\mathcal{L}$ in one dimension
is that it changes behavior near $R_M=0$ because
the interference term (linear in $t$ in Equation~\ref{eq:tdratemult})
becomes unconstrained.  Therefore,
we estimate the consistency of the data with no mixing
using a frequentist method.  Generating 1000 simulated
data sets with no mixing,
each with 76,300 events representing signal
and background in the quantities
$\{\mKppp,\dm,\tKppp\}$, we find 4.3\% of 
simulated data sets have a fitted value of $R_M$
greater than that in the observed data set.
We conclude that the observed data are consistent
with no mixing at the 4.3\% confidence level.

\begin{figure}[!htp]
\begin{center}
\includegraphics[width= 0.45\linewidth]{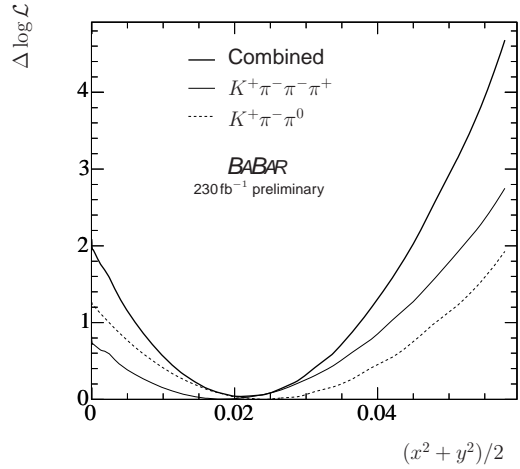}
\caption[Contours of $\Delta\ln\mathcal{L}$ levels for combined result.]
{
$\Delta\ln\mathcal{L}$ as a function of $R_M$ for separate
and combined results of $K^+ \pi^- \pi^+ \pi^-$ and $K^+ \pi^- \pi^0$.
(On the axis, the natural logarithm is denoted $\log$.)
}
\label{fig:nll_rmix_combine}
\end{center}
\end{figure}

We combine the value of $R_M$ from this analysis with that obtained from an
analysis of the
decays $\Dz \to K^+ \pi^- \pi^0$~\cite{Wilson:2006sj}
by adding the $\Delta\ln\mathcal{L}(R_M)$ curves from the
two separate analyses.  The $\Delta\ln\mathcal{L}(R_M)$ curves are shown in
Fig~\ref{fig:nll_rmix_combine}.
We extract a central value and an uncertainty from the
combined curve using the same procedure as for each individual result.
With this method, we find $R_M = (0.020\,\mbox{}^{+0.011}_{-0.010})\%$,
where the uncertainty is statistical only.
We determine the upper limit $R_M < 0.042\%$ at the 95\% confidence level,
and we find
the combined data are consistent with the no-mixing hypothesis
at the 2.1\% confidence level,
as determined from the $\Delta\ln\mathcal{L}(R_M)$ curve.

\section{CONCLUSION}
\label{sec:Summary}

\begin{sloppypar}
We find that the data used in an
analysis of \wsdecay\ are consistent with the no-mixing hypothesis
at the 4.3\% confidence level.
Assuming \CP\ conservation, we measure the time-integrated mixing rate
$R_M = (0.019\ \mbox{}^{+0.016}_{-0.015}\,(\textrm{stat.}) \pm 0.002\,(\textrm{syst.})) \%$,
and $R_M < 0.048\%$ at the 95\% confidence level.
Furthermore, we combine these results with those of a similar analysis
of the decays $\Dz \to K^+ \pi^- \pi^0$~\cite{Wilson:2006sj}.  From this combination,
we find $R_M = (0.020\,\mbox{}^{+0.011}_{-0.010}\,(\textrm{stat.}))\%$
and $R_M < 0.042\%$ at the 95\% confidence level.  The combined data sets
are consistent with the no-mixing hypothesis with 2.1\% confidence.
\end{sloppypar}

\section{ACKNOWLEDGMENTS}
\label{sec:Acknowledgments}
We are grateful for the 
extraordinary contributions of our \pep2\ colleagues in
achieving the excellent luminosity and machine conditions
that have made this work possible.
The success of this project also relies critically on the 
expertise and dedication of the computing organizations that 
support \babar.
The collaborating institutions wish to thank 
SLAC for its support and the kind hospitality extended to them. 
This work is supported by the
US Department of Energy
and National Science Foundation, the
Natural Sciences and Engineering Research Council (Canada),
Institute of High Energy Physics (China), the
Commissariat \`a l'Energie Atomique and
Institut National de Physique Nucl\'eaire et de Physique des Particules
(France), the
Bundesministerium f\"ur Bildung und Forschung and
Deutsche Forschungsgemeinschaft
(Germany), the
Istituto Nazionale di Fisica Nucleare (Italy),
the Foundation for Fundamental Research on Matter (The Netherlands),
the Research Council of Norway, the
Ministry of Science and Technology of the Russian Federation, 
Ministerio de Educaci\'on y Ciencia (Spain), and the
Particle Physics and Astronomy Research Council (United Kingdom). 
Individuals have received support from 
the Marie-Curie IEF program (European Union) and
the A. P. Sloan Foundation.

\end{document}